\documentclass[prd,twocolumn,amsmath,amssymb,nofootinbib]{revtex4-1}

\usepackage{latexsym,ifthen,graphics,color,epsfig,hyperref,mathrsfs}

\newcommand{\ie}{{i.e.}}

\newcommand{\wrt}{with respect to}
\newcommand{\lhs}{left-hand side}

\newcommand{\be}{\begin{equation}}
\newcommand{\ee}{\end{equation}}
\newcommand{\bea}{\begin{eqnarray}}
\newcommand{\eea}{\end{eqnarray}}
\newcommand{\beas}{\begin{eqnarray*}}
\newcommand{\eeas}{\end{eqnarray*}}

\newcommand{\bear}{\begin{array}{l}}
\newcommand{\eear}{\end{array}}

\newcommand{\bcf}{\begin{center}\begin{figure}}
\newcommand{\ecf}{\end{figure}\end{center}}

\newcommand{\bct}{\begin{center}\begin{table}}
\newcommand{\ect}{\end{table}\end{center}}

\newcommand{\eq}[1]{(\ref{eq:#1})}

\newcommand{\eqs}[2]{(\ref{eq:#1}) and~(\ref{eq:#2})}

\newcommand{\sect}[1]{section~\ref{sec:#1}}

\newcommand{\app}[1]{appendix~\ref{app:#1}}

\newcommand{\D}{d}

\newcommand{\Int}[1]{\int \!\! d^{\D} \! #1 \,}

\newcommand{\MomInt}[2]{\int \!\! \frac{d^{#1} #2}{(2\pi)^{#1}} \, }

\newcommand{\Fint}[1]{\int \mathcal{D} #1 \,}

\newcommand{\der}[2]{\frac{d #1}{d #2}}
\newcommand{\pder}[2]{\frac{\partial #1}{\partial #2}}

\newcommand{\fder}[2]{\frac{\delta #1}{\delta #2}}
\newcommand{\dfder}[3]{\frac{\delta^2 #1}{\delta #2 \delta #3}}

\newcommand{\hf}{\frac{1}{2}}

\newcommand{\Tr}{\mathrm{Tr}\,}

\newcommand{\deltahat}[1]{
	\bar{\delta}(#1)
}

\newcommand{\field}{\phi}

\newcommand{\cutoff}{K}

\newcommand{\ep}{C}

\newcommand{\dd}{\dot{\ep}}
\newcommand{\knl}[1]{\cdot {#1}\cdot}

\newcommand{\flow}{\Lambda \partial_\Lambda}
\newcommand{\totalflow}[1]{\Lambda \der{#1§}{\Lambda}}

\newcommand{\Stot}{S}
\newcommand{\Sint}{\mathcal{S}}

\newcommand{\dil}[1]{\mathrm{D}^{(#1)}}

\newcommand{\classical}[3]{\fder{#1}{\field} \knl{#2} \fder{#3}{\field}} 
\newcommand{\quantum}[2]{\fder{}{\field} \knl{#1} \fder{#2}{\field}}

\newcommand{\hepth}[1]{hep-th/#1}
\newcommand{\hepph}[1]{hep-ph/#1}

\newcommand{\condmat}[1]{cond-mat/#1}

\begin{document}

\title{Relationships Between Exact RGs and some Comments on Asymptotic Safety}

\author{Oliver~J.~Rosten}

\affiliation{Department of Physics and Astronomy, University of Sussex, Brighton, BN1 9QH, U.K.}
\email{O.J.Rosten@Sussex.ac.uk}



\begin{abstract}
	The standard flow equation for the effective average action can be derived from a
	Legendre transform of Polchinski's
	exact renormalization group equation. However, the latter is not well adapted for finding
	fixed-points with non-zero anomalous dimension. Instead, it is more convenient to use a 
	modified version which ensures that the redundant coupling associated with the normalization
	of the field never appears in the action. Taking this as the starting point, a Legendre transform
	is constructed allowing a direct derivation of the corresponding flow equation for the 
	effective average action. This equation is then used to exactly construct some illuminating
	(though essentially trivial) asymptotically safe trajectories emanating from various
	non-unitary fixed-points. 
	Finally, in the context of asympotically safe quantum gravity, it
	is pointed out that the standard argument that the
	anomalous dimension of Newton's constant is necessarily $2-\D$ at a non-trivial fixed-point is incomplete.
	The implications of this are discussed.
\end{abstract}

\maketitle

\section{Introduction}

Superficially, there are two genera of Exact Renormalization Group (ERG) equations: those involving the Wilsonian effective action, $\Stot$, and those involving the effective average action, $\Gamma$. However, as we will make clear in this letter, a more apt analogy is that certain representatives of these two sets are better classified as identical twins: they contain precisely the same (genetic) information but, as is typical of such twins, nevertheless have distinct appearances and personalities.

Of course, there is already a well known example of this. The standard flow equation for the effective average action~\cite{Wetterich-1PI,TRM-ApproxSolns,Ellwanger-1PI,Bonini-1PI} is related to Polchinski's equation~\cite{Pol} by a Legendre transform. Below we will show how this relationship carries over when each of these equations is tweaked to conveniently take account of the anomalous dimension of the field.

Having done this, we will use the effective average action formalism to uncover an infinite number of asymptotically safe RG trajectories in scalar field theory. Whilst these are trivial in the sense that the effective action has only a two-point contribution, the fact that everything can be done exactly is illuminating. In particular, every last one of these trajectories is unphysical, since the corresponding ultraviolet (UV) theory is non-unitary. This serves as an important lesson for asymptotic safety scenarios in general.

With this in mind, we revisit the issue of the `anomalous dimension' associated with Newton's constant in asymptotically safe quantum gravity.
Due to the structure of the flow equation, which depends separately on the background metric and a fluctuation, it turns out that the anomalous dimension of the latter is \emph{not} necessarily equal to $2-\D$ (which would be the value deduced  purely on scaling grounds~\cite{Reuter+Lauscher-HD}) at a non-trivial fixed-point.\footnote{This is in direct analogy with the fact that, needless to say, fixed-points in scalar field theory do not necessarily have an anomalous dimension of zero.} This opens up the possibility of a richer spectrum of fixed-points than has been hitherto found, though the crucial question as to their unitarity remains unanswered.

\section{Relationships between Flow Equations}

In the case of the Wilsonian effective action, various flow equations (\ie\ species of the ERG) follow from different choices of the (continuum version of the) blocking procedure. Specifically, let us suppose that degrees of freedom are coarse-grained over patches of characteristic size $1/\Lambda$. Then we can write the effective field, $\field$, in terms of the bare field, $\field_0$, as
\be
	\phi(x) = b_\Lambda[\phi_0](x).
\ee
An obvious choice to make would be $b_\Lambda[\phi_0](x) = \int_y f(x-y;\Lambda)\phi_0(y)$, where (quasi-) locality is implemented by demanding that $f(z;\Lambda$) decays rapidly for $z\Lambda >1$. However, there are many other perfectly valid choices of $b$ and, indeed, there is no need for the blocking procedure to be linear in the field. Whatever we choose for $b$, the Wilsonian effective action can be related to the bare action via
\be
	e^{-\Stot_\Lambda[\phi]} = \Fint{\phi_0} \delta\bigl[ \phi - b_\Lambda[\phi_0]\bigr]
	e^{-\Stot_{\Lambda_0}[\phi_0]},
\label{eq:blocking}
\ee
where $\Lambda_0$ is the bare scale.
ERG equations follow from differentiating \wrt\ $\Lambda$ at constant field. Defining an object $\Psi$ via~\cite{mgierg1}
\be
	\Psi(x) e^{-\Stot_\Lambda[\phi]}
	=
	 \Fint{\phi_0} \delta\bigl[ \phi - b_\Lambda[\phi_0]\bigr]
	 \Lambda \pder{b_\Lambda[\phi_0](x)}{\Lambda}
	e^{-\Stot_{\Lambda_0}[\phi_0]},
\label{eq:Psi-explicit}
\ee
it follows that
\be
-\flow e^{-\Stot_\Lambda[\field]} =  \Int{x} \fder{}{\field(x)} 
	\left\{
	\Psi(x) e^{-\Stot_\Lambda[\field]}
	\right\}.
\label{eq:blocked}
\ee
Different choices of $\Psi$ (equivalently different blocking procedures) yield different ERG equations. Note that~\eq{blocked} can also be derived by considering the effects of an infinitesimal field redefinition
$\field \rightarrow \field - \delta\Lambda /\Lambda \, \Psi$, as first recognized by Wegner~\cite{WegnerInv} and later explored by Latorre and Morris~\cite{TRM+JL}.

In preparation for giving the explicit form of Polchinski's equation, let us introduce a (momentum space) ultraviolet (UV) cutoff function, $\cutoff(p^2/\Lambda^2)$, satisfying the following properties: (i) it is quasi-local meaning that, for small momentum, it is analytic in $p^2/\Lambda^2$, (ii) for $p^2/\Lambda^2 \gtrsim 1$ it dies off faster than any power~\cite{Fundamentals}, (iii) its derivative is negative definite~\cite{HO-Remarks}.%
\footnote{%
This follows because Polchinski's equation, when written in terms of
$e^{-\Sint}$, takes the form of a heat equation. Therefore, for the evolution with decreasing $\Lambda$ to correspond to a well-posed problem, given a generic starting point, we require that $d \cutoff(x) /d x <0$.
}
Defining
\be
	\ep_\Lambda(p^2) \equiv \frac{\cutoff(p^2/\Lambda^2)}{p^2},
\ee
it is convenient to split up the action in the following manner:
\be
	\Stot_\Lambda[\field] = \hf \field \cdot \ep^{-1}_\Lambda \cdot \field + \Sint_\Lambda[\field],
\label{eq:split}
\ee
where $\field \cdot \ep^{-1}_\Lambda \cdot \field = \MomInt{\D}{p} \field(-p) \ep^{-1}_\Lambda(p^2) \field(p)$.
Defining $\dot{X} \equiv -\Lambda d X/d\Lambda$, Polchinski's equation follows from taking
\be
	\Psi(p) = \hf \dd_\Lambda(p^2) 
	\biggl\{
		\fder{\Sint_\Lambda[\field]}{\field(-p)} - \ep^{-1}_\Lambda(p^2) \field(p)
	\biggr\},
\label{eq:choice}
\ee
which yields the flow equation
\be
	-\flow \Sint_\Lambda[\field] = \hf \classical{\Sint}{\dd_\Lambda}{\Sint} - \hf \quantum{\dd_\Lambda}{\Sint}.
\label{eq:Pol}
\ee

There are a few different ways to derive the flow equation for the effective average action from the Polchinski equation~\cite{TRM-ApproxSolns,Ellwanger-1PI,OJR-1PI,HO-Remarks}. Three of these approaches~\cite{TRM-ApproxSolns,Ellwanger-1PI,OJR-1PI} share a common feature: 
by introducing a source, an appropriately infrared (IR) regularized generator of connected Green's functions can be defined and, from this, the effective average action constructed via a Legendre transform.
Rather than following this route directly, we will exploit the fact that this strategy induces a Legendre transform map between the effective average action and the Wilsonian effective action~\cite{TRM-ApproxSolns}, which will be our starting point:
\be
	\Gamma_\Lambda[\Phi] = \Sint_\Lambda[\field]
	-
	\hf
	\bigl(
		\Phi  -\field
	\bigr)
	\cdot
		D^{\Lambda_0}_\Lambda
	\cdot
	\bigl(
		\Phi  - \field
	\bigr),
\label{eq:map-Pol}
\ee
where
\be
	D^{\Lambda_0}_\Lambda(p^2) = \frac{p^2}{\cutoff(p^2/\Lambda_0^2) - \cutoff(p^2/\Lambda^2)}.
\ee
The flow equation satisfied by $\Gamma_\Lambda$ can be deduced as follows. First of all, note that
\begin{subequations}
\begin{align}
	\fder{\Sint_\Lambda[\field]}{\field(p)} &= -D^{\Lambda_0}_\Lambda(p^2)
	\bigl[
		\Phi(-p)  - \field(-p)
	\bigr],
\label{eq:S_phi}
\\
	\fder{\Gamma_\Lambda[\Phi]}{\Phi(p)} & = 
	- D^{\Lambda_0}_\Lambda(p^2)
	\bigl[
		\Phi(-p)  - \field(-p)
	\bigr].
\label{eq:Gamma_Phi}
\end{align}
\end{subequations}
Using the latter result it is easy to check that
\be
	\flow \bigr\vert_\field \Sint_\Lambda[\field] =
	\flow \bigr\vert_\Phi \Gamma_\Lambda[\Phi] 
	- \hf 
	\bigl(
		\Phi  -\field
	\bigr)
	\cdot
		\dot{D}^{\Lambda_0}_\Lambda
	\cdot
	\bigl(
		\Phi  - \field
	\bigr)
\label{eq:flows}
\ee
where, as before, an overdot stands for $\Lambda d / d\Lambda$. Noting that $\dot{D}^{\Lambda_0}_\Lambda = D^{\Lambda_0}_\Lambda \dot{\ep}_\Lambda D^{\Lambda_0}_\Lambda $, the Polchinski equation can be recast as
\be
	-\flow \Bigr\vert_\Phi \Gamma_\Lambda[\Phi] = - \hf \quantum{\dd_\Lambda}{\Sint}.
\ee
From~\eqs{S_phi}{Gamma_Phi} it is apparent that
\begin{multline}
	\MomInt{\D}{q}
	\biggl\{
		\dfder{S_\Lambda[\field]}{\field(p)}{\field(q)} - D^{\Lambda_0}_\Lambda(p^2) \deltahat{p+q}
	\biggr\}
\\
	\times
	\biggl\{
		\dfder{\Gamma_\Lambda[\field]}{\Phi(-q)}{\Phi(-p')} + D^{\Lambda_0}_\Lambda(q^2) \deltahat{p'+q}
	\biggr\}
	=
\\
	-\bigl[D^{\Lambda_0}_\Lambda(p^2)\bigr]^2
	\deltahat{p-p'},
\end{multline}
where $\deltahat{q} \equiv (2\pi)^{\D} \delta^{(\D)}(q)$. Defining
$\Gamma^{(2)}_\Lambda = \delta^2 \Gamma_\Lambda / \delta \Phi \delta \Phi$ and discarding a vacuum energy term, we arrive at the standard flow equation for the effective average action:
\be
	-\flow \Gamma_\Lambda[\Phi]
	=
	\hf \Tr
	\Bigl\{
	 \dot{D}^{\Lambda_0}_\Lambda
	 \Bigl[
	 	D^{\Lambda_0}_\Lambda + \Gamma_\Lambda^{(2)}
	 \Bigr]^{-1}
	 \Bigr\}.
\label{eq:Morris-1PI}
\ee
Note that Wetterich's form of this equation~\cite{Wetterich-1PI} contains a minor difference: it can be obtained by shifting $\Gamma_\Lambda[\Phi] \rightarrow \Gamma_\Lambda[\Phi] + \hf \int_p \Phi(p) \Phi(-p) p^2$. 

It is worth commenting on the limit $\Lambda_0 \rightarrow \infty$. As far as either $\Sint$ or $\Gamma$ is concerned, the existence of this limit presumes that the theory
sits either at a fixed-point or on a renormalized trajectory (which arises from perturbing a fixed-point in its relevant directions~\cite{Wilson,TRM-Elements}). Nevertheless, even if we are dealing with a non-renormalizable theory, it is interesting to note that if we make the \emph{replacement} $D_\Lambda^{\Lambda_0} \rightarrow D_\Lambda^{\infty}$ in~\eq{Morris-1PI} then the flow equation remains UV regularized. This is as a consequence of the appearance of $ \dot{D}^{\infty}_\Lambda$, which decays rapidly in the UV.
However, by making this replacement (rather than taking the limit for all ingredients of the flow equation) we are changing our definition of $\Gamma$. Calling this new object $\Gamma'$, and making the bare scale explicit, it is related to the Wilsonian effective action via
\be
	\Gamma'^{\Lambda_0}_{\,\Lambda}[\Phi] = \Sint^{\Lambda_0}_\Lambda[\field]
	-
	\hf
	\bigl(
		\Phi  -\field
	\bigr)
	\cdot
		D^{\infty}_\Lambda
	\cdot
	\bigl(
		\Phi  - \field
	\bigr)
.
\label{eq:map-Pol-mod}
\ee
Since $\Gamma_\Lambda$ generates IR regularized correlation functions, the same cannot be true of $\Gamma'$. Nevertheless, for flows which pass close to an IR fixed-point, the differences between the vertices of 
$\Gamma$ and $\Gamma'$ will be of order $\mathrm{momenta}^2/\Lambda^2_0$. Moreover, by combining~\eqs{map-Pol}{map-Pol-mod} it is possible to reconstruct $\Gamma$ from $\Gamma'$, though one should bear in mind that, for most practical purposes, this will be complicated by the necessity to truncate.

As we stated at the beginning of this analysis, \eq{map-Pol} follows as a result of defining $\Gamma$ by introducing an IR regularized generator of connected correlation functions. However, as in~\cite{HO-Remarks}, one can take a different view and simply define $\Gamma$ via an equation of the form
\be
	\Gamma_\Lambda[\Phi] = S_\Lambda[\field] 
	+ \field \cdot \mathcal{P}_\Lambda^{\Lambda_0} \cdot \Phi
	-\hf \field \cdot \mathcal{Q}_\Lambda^{\Lambda_0} \cdot \field
	- \hf \Phi \cdot \mathcal{R}_\Lambda^{\Lambda_0} \cdot \Phi,
\label{eq:Legendre-Guess}
\ee
where now $\Phi$ is \emph{defined} via the generalization of~\eq{S_phi} 
\be
	\fder{\Sint_\Lambda[\field]}{\field(p)} = \mathcal{Q}_\Lambda^{\Lambda_0}(p^2) \field(-p) 
	- \mathcal{P}_\Lambda^{\Lambda_0}(p^2) \Phi(-p)
\label{eq:Phi-def}
\ee
and we fix $\mathcal{P}$, $\mathcal{Q}$ and $\mathcal{R}$
by demanding that the flow equation satisfied by $\Gamma$ takes the form~\eq{Morris-1PI}.
Of course, in the current scenario we know the result of this procedure: we find
that $\mathcal{P}_\Lambda^{\Lambda_0}(p^2) = \mathcal{Q}_\Lambda^{\Lambda_0}(p^2) = \mathcal{R}_\Lambda^{\Lambda_0}(p^2) = D^{\Lambda_0}_\Lambda(p^2)$. However, this method provides a neat way~\cite{HO-Remarks} of deriving the flow equation satisfied by $\Gamma$
in the case that the Polchinski equation is deformed to allow for a convenient treatment of the anomalous dimension of the field.

The particular deformation of the Polchinski equation that we will consider follows from shifting
\be
	\Psi \rightarrow \Psi - \frac{\eta}{2} \field.
\label{eq:shift}
\ee
This extra field redefinition is chosen such that the field strength renormalization---which is a redundant (equivalently inessential) coupling---is removed from the action; this is achieved by identifying, as usual, $\eta = \Lambda \,d\ln Z/d\Lambda$. At a given critical fixed-point, $\eta$ obtains a universal value $\eta_\star$; the allowed values of $\eta_\star$ are discrete~\cite{Fundamentals} and the fixed-point flow equation can be thought of as a non-linear eigenvalue equation for $\eta_\star$~\cite{TRM-Elements}. Away from a fixed-point, $\eta$ is non-universal and can be fixed by a renormalization condition (say that the kinetic term is canonically normalized along the flow) or by appealing to global aspects of the flow, as we shall see later. In fact, so long as one considers flows in the vicinity of a given fixed-point, it is acceptable to take $\eta(\Lambda) = \eta_\star$; indeed, in this case it has been shown how to readily derive the flow equation for the effective average action starting from~\eq{Legendre-Guess}~\cite{HO-Remarks}. However, if one wishes to allow for the possibility of flows between different fixed-points, then $\eta$ must be allowed to depend on the scale. In what follows, we shall treat this case.

The shift~\eq{shift} yields the flow equation of~\cite{Ball}:
\begin{multline}
	\Bigl(-\flow +\frac{\eta}{2} \field \cdot \fder{}{\field}\Bigr)
	\Sint_\Lambda[\field] = \hf \classical{\Sint}{\dd_\Lambda}{\Sint} 
\\
	- \hf \quantum{\dd_\Lambda}{\Sint}
	-\frac{\eta}{2} \field \cdot \ep_\Lambda^{-1} \cdot \field;
\label{eq:Ball}
\end{multline}
the aim now is to choose $\mathcal{P}$, $\mathcal{Q}$ and $\mathcal{R}$ such that the effective average action satisfies
\be
	\Bigl(-\flow +\frac{\eta}{2} \Phi \cdot \fder{}{\Phi}\Bigr) \Gamma_\Lambda[\Phi]
	=
	\hf \Tr
	\biggl[
		f_\Lambda^{\Lambda_0}
		\Bigl(
			F_\Lambda^{\Lambda_0}+ \Gamma_\Lambda^{(2)}
		\Bigr)^{-1}
	\biggr],
\label{eq:Gamma-Flow}
\ee
where $f_\Lambda^{\Lambda_0}(p^2)$ and $F_\Lambda^{\Lambda_0}(p^2)$ will shortly be determined in terms of the cutoff function, $\cutoff$, and the anomalous dimension.

Defining
\be
	\zeta_\Lambda^{\Lambda_0} \equiv \int^{\Lambda_0}_\Lambda \frac{d\Lambda'}{\Lambda'} \eta(\Lambda'),
\label{eq:gamma}
\ee
let us introduce the function
\begin{multline}
	\sigma_\Lambda^{\Lambda_0}(p^2) \equiv \cutoff(p^2/\Lambda^2)\, e^{\zeta_\Lambda^{\Lambda_0}} 
\\
	\times
	\biggl[
	a(p^2/\Lambda_0^2)-
	\int^{\Lambda_0}_\Lambda d\Lambda' e^{-\zeta_{\Lambda'}^{\Lambda_0}} 
	\der{}{\Lambda'} \frac{1}{\cutoff(p^2/\Lambda'^2)}
	\biggr],
\label{eq:sigma}
\end{multline}
where $a$ is a constant of integration. As we will describe in more detail in
\app{Details}, we can go from~\eq{Ball} to~\eq{Gamma-Flow} so long as we 
take
\begin{subequations}
\begin{align}
	\mathcal{P}_\Lambda^{\Lambda_0}(p^2) & = \frac{ p^2}{\sigma_\Lambda^{\Lambda_0}(p^2)}
	b(p^2/\Lambda_0^2),
\label{eq:P}
\\
	\mathcal{Q}_\Lambda^{\Lambda_0}(p^2) & = \frac{p^2}{\cutoff(p^2/\Lambda^2)}
	\biggl[
		\frac{1}{\sigma_\Lambda^{\Lambda_0}(p^2)} - 1
	\biggr],
\label{eq:Q}
\\
	\mathcal{R}_\Lambda^{\Lambda_0}(p^2) & = 
	p^2
	\biggl[
	\frac{\cutoff(p^2/\Lambda^2)}{\sigma_\Lambda^{\Lambda_0}(p^2)} b^2(p^2/\Lambda_0^2)
	+ c(p^2/\Lambda_0^2) e^{-\zeta_\Lambda^{\Lambda_0}}
	\biggr]
\label{eq:R},
\end{align}
\end{subequations}
(where $b$ and $c$ are further constants of integration) and identify
\be
	F_\Lambda^{\Lambda_0}(p^2) = \mathcal{R}_\Lambda^{\Lambda_0}(p^2),
	\qquad
	f_\Lambda^{\Lambda_0}(p^2) = \bigl[\mathcal{P}_\Lambda^{\Lambda_0}(p^2)\bigr]^2 \dd_\Lambda(p^2).
\label{eq:f+F}
\ee
Notice that we can write
\be
	f = -e^{-\zeta} \totalflow{}Fe^\zeta.
\label{eq:fandF}
\ee

The three constants of integration can be fixed by demanding that 
$\mathcal{P}$, $\mathcal{Q}$ and $\mathcal{R}$ are all quasi-local and
all remain both finite and non-zero in the limit
$\Lambda_0 \rightarrow \infty$. Notice that since $\mathcal{P}$, $\mathcal{Q}$ and $\mathcal{R}$ all
involve $\eta(\Lambda_0)$, the existence of the limit at the level of these objects, as opposed to at
the level of the action, superficially appears to require
that we are dealing with a renormalizable theory. This is apparently in distinction to the
corresponding analysis for~\eq{Morris-1PI} for which we recall that
there was no complication in defining $D^{\infty}_\Lambda$.

However, for non-renormalizable theories, since we do not hit a fixed-point in the UV limit (indeed, the limit doesn't exist at the level of the action) we are free to choose the UV behaviour of $\eta$ to be whatever we like. 
(This should become clearer from the perspective of \sect{Examples}). Therefore, there is nothing to stop us from defining a function $\eta(\Lambda)$ which exists for $\infty \geq \Lambda \geq 0$ and then making the replacements $f_\Lambda^{\Lambda_0} \rightarrow f_\Lambda^{\infty}, \  F_\Lambda^{\Lambda_0} \rightarrow 
F_\Lambda^{\infty}$ in~\eq{Gamma-Flow}, should we so desire. 
Indeed, we are free to do  this even if the action is defined only for $\Lambda_0 < \infty$ or, worse still,
blows up at some finite scale due to the existence of a Landau pole. For example, in the case that the flow hits an \emph{IR} fixed-point, a perfectly legitimate choice would be $\eta(\Lambda) = \eta_\star^{\mathrm{IR}}, \forall \;\Lambda$~\cite{HO-Remarks}, irrespective of any pathological behaviour of the action in the UV.
Of course, if we make the aforementioned replacements in the flow equation then we change the definition and interpretation of $\Gamma$, just as we did above.

Our strategy, then, is as follows. Bearing in mind that the limit $\Lambda_0 \rightarrow \infty$ \emph{must} exist for 
renormalizable theories, we will determine $a$, $b$ and $c$ by working in this context. We are, of course, at liberty to take different choices for non-renormalizable theories, since here there is no absolute requirement for 
$\mathcal{P}$, $\mathcal{Q}$ and $\mathcal{R}$ to exhibit a limit as $\Lambda_0 \rightarrow \infty$. Equally, though, we are perfectly entitled to stick with our choices for renormalizable theories and, by doing so, this has the benefit of allowing us the option of making the above replacements in the flow equation. 

For renormalizable theories, 
$\eta(\Lambda \rightarrow \infty) = \eta_\star$. Moreover, if we take $\Lambda$ to be sufficiently large
then $\eta$ will be approximately constant over the range of integration in~\eq{gamma}.
With this in mind, focusing on the second contribution to $\sigma$ it is easy to check that, in the fixed-point regime,
we can only send $\Lambda_0 \rightarrow \infty$ 
if $\eta_\star <2$. But this restriction is no surprise, since it 
has a physical origin: it is necessary for the corresponding fixed-point
to be critical (for an explanation of this in the context of the ERG, see~\cite{Fundamentals}).
For $\eta_\star\geq 2$, fixed-points can exist but are non-critical.
As such, they exhibit zero correlation length and are sinks of RG trajectories~\cite{Wegner_CS}.
Therefore, fixed-points with $\eta_\star \geq 2$ are always IR fixed-points
and so it is no surprise that they cannot appear in the UV limit $\Lambda_0 \rightarrow \infty$.
So, henceforth, in the regime of large $\Lambda$ we restrict ourselves to $\eta<2$.

With this in mind, we now turn to the first contribution to $\sigma$
which, in the fixed-point regime,
behaves like
\[
	\cutoff(p^2/\Lambda^2) e^{\zeta_\Lambda^{\Lambda_0}} a(p^2/\Lambda_0^2)
	\approx \cutoff(p^2/\Lambda^2) (\Lambda_0/\Lambda)^{\eta_\star} a(p^2/\Lambda_0^2).
\]
Let us start by supposing that $2>\eta_\star >0$. In this case, there is clearly
a divergent contribution. Whilst this could be compensated for
by adjusting $a$, this would (given the allowed range of $\eta_\star$) 
necessarily result in a loss of quasi-locality. Therefore we
conclude that $a=0$. 

Sticking with $2>\eta_\star >0$, note that 
there is no need to set either $b$ or $c$ to zero: $b$ multiplies things which
are finite in the $\Lambda_0 \rightarrow \infty$ limit, whereas $c$ multiplies something
which vanishes. We can fix $b$ and $c$ by extending our analysis to $\eta =0$
and
comparing with
the plain Polchinski equation, where we know that $\mathcal{P} = \mathcal{Q} = \mathcal{R} = 
D^{\Lambda_0}_\Lambda$. It is easy to check that $a=0$ is already compatible
with this, and that we should take $b = c = 1/\cutoff(p^2/\Lambda_0^2)$.

For $\eta_\star <0$, the situation is rather different. In this case, the
`$c$' term contributing to $\mathcal{R}$ now contains a divergent piece 
in the limit $\Lambda_0 \rightarrow \infty$. For $0>\eta_\star>-2$, at any rate,
quasi-locality means that we are compelled to set $c=0$. This discontinuity
between $\eta_\star\geq0$ and $\eta_\star<0$ is interesting since theories
of the latter type are expected to be non-unitary upon continuation to
Minkowski space~\cite{Mack-Unitary,WeinbergI}; indeed, it is tempting to speculate that we are seeing a
manifestation of this. For $\eta_\star = -2$ (for which a non-unitary fixed-point
most certainly exists---see below and also~\cite{Wegner_CS,Fundamentals}), quasi-locality can be maintained
whilst taking a non-zero $c$, but there seems no need to introduce
further discontinuities and so we shall stick with $c=0$ for $\eta_\star <0$.
There are no restrictions on $a$ and $b$ besides quasi-locality and so
it is natural to fix them to what we obtained for $\eta_\star \geq 0$.
To summarise, a sensible choice of integration constants is:
\be
\begin{split}
	a = 0,
	\qquad
	b = \cutoff^{-1}(p^2/\Lambda_0^2),
\\
	c=
	\biggl\{
	\begin{matrix}
		\cutoff^{-1}(p^2/\Lambda_0^2), & 0 \leq \eta(\Lambda_0) < 2
	\\
		0, & \eta(\Lambda_0) < 0.
	\end{matrix}
\end{split}
\ee

Let us make a few comments on the flow equation before moving on. The flow equation~\eq{Gamma-Flow}
was written down by Morris in~\cite{TRM-Deriv}. The relationship between
$f$ and $F$, given by \eq{fandF}, was deduced using general (though heuristic) arguments. In
this letter, we have shown (building on~\cite{HO-Remarks}) how this flow
equation follows directly from the underlying Wilsonian formalism. (In~\cite{OJR-1PI}, a different method was
used to derive the fixed-point version of~\eq{Gamma-Flow}.)
As claimed,
then, the pair of equations~\eqs{Ball}{Gamma-Flow} have the same genotype
but rather different phenotypes.

For applications such as finding fixed-points, it is often useful to work in dimensionless variables. To this end, we define
\be
	\tilde{x} \equiv x\Lambda, \qquad \tilde{p} \equiv p/\Lambda
\label{eq:Rescale-Mom}
\ee 
and take
\be
	\tilde{\Phi}(\tilde{x}) = \Phi(x) \Lambda^{-(\D-2)/2},
	\qquad
	\tilde{\Phi}(\tilde{p}) = \Phi(p) \Lambda^{(\D+2)/2}
.
\label{eq:Rescale-Fields}
\ee
Introducing the RG time $t \sim -\ln \Lambda$ we take $\tilde{\Gamma}_t[\tilde{\Phi}] = \Gamma_\Lambda[\Phi]$, $\tilde{f}^{t_0}_t(\tilde{p}) = f_\Lambda^{\Lambda_0} (p^2) /\Lambda^2$ and $\tilde{F}^{t_0}_t(\tilde{p}) = f_\Lambda^{\Lambda_0}  (p^2) /\Lambda^2$. Now the flow equation can be written as
\be
	\Bigl(\partial_t +\dil{\delta} \tilde{\Phi} \cdot \fder{}{\tilde{\Phi}}\Bigr) \tilde{\Gamma}_t[\tilde{\Phi}]
	=
	\hf \Tr
	\biggl[
		\tilde{f}^{t_0}_t
		\Bigl(
			\tilde{F}^{t_0}_t+ \tilde{\Gamma}_t^{(2)}
		\Bigr)^{-1}
	\biggr],
\label{eq:Gamma-Flow-Dim}
\ee
where $\delta = [\D-2+\eta(t)]/2$ and $\dil{\delta}$ is the dilatation generator so that, in momentum space,
we have
\be
	\dil{\delta} \tilde{\Phi} (\tilde{p}) = (\delta -\D - \tilde{p} \cdot \partial_{\tilde{p}}) \tilde{\Phi}(\tilde{p})
.
\ee
Henceforth, we will drop all tildes. With this in mind, fixed-points are defined according $\partial_t \Gamma_\star[\Phi] = 0$, with $\eta(t) = \eta_\star$, where we understand that $f^{t_0}_t \rightarrow f^{-\infty}_t$
and $F^{t_0}_t \rightarrow F^{-\infty}_t$.

\section{Some Examples of Asymptotic Safety}
\label{sec:Examples}

Solutions of~\eq{Gamma-Flow-Dim} for which $\Gamma_t[\Phi]$ does not have any contributions beyond the two-point level are trivial to find. Indeed, for non-gravitational theories we can ignore vacuum terms [which has anyway been done in our derivation of~\eq{Gamma-Flow-Dim}], so we can take
\be
	\Gamma_t[\Phi] = \hf \Phi \cdot h_t \cdot \Phi.
\ee
Substituting this into~\eq{Gamma-Flow-Dim} we find that%
\footnote{Strictly, we are abusing notation since $\eta(t)$ is not the same function of its argument as $\eta(\Lambda)$. But what we mean should be clear from the context.
}
\be
	\bigl[ \partial_t  + 2 p^2 \partial_{p^2} + \eta(t)  -2 \bigr] h_t(p^2) = 0
.
\ee

The Gaussian fixed-point corresponds to
\be
	h_\star(p^2) = a_0 p^2,	\quad \eta_\star = 0
\ee
where $a_0>0$ is a free parameter related to the normalization of the field (see~\cite{Fundamentals,OJR-1PI,HO-Remarks} for a much more detailed discussion of this). There is also an infinite set of non-unitary fixed-points with $\eta_\star <0$, first discovered by Wegner~\cite{Wegner_CS}:
\be
	h_\star(p^2) = a_{2n} p^{2(n+1)}, \quad \eta_\star = -2n,
	\quad n = 1,2,\ldots
\ee

We will now show that there exist trajectories emanating from these fixed-points which end at the Gaussian fixed-point in the IR (it will also become apparent that there are trajectories between the various non-unitary fixed-points). To see this, note that the general solution for $h_t$ is
\be
	h_t(p^2) = p^2 e^{-\int_{t_0}^t dt' \eta(t')} w_{t_0} \bigl(p^2 e^{-2(t-t_0)}\bigr), 
\ee
where $t_0$ is the bare scale and $w$ is a quasi-local, but otherwise arbitrary, function. As a first example, we will consider a flow between the fixed-point with $\eta_\star = -2$ and the Gaussian fixed-point. For $\eta(t)$ we will (to start with) take the following ansatz:
\be
	\eta(t) = 
	\Biggl\{
	\begin{matrix}
		-2 & \quad t<t_r
	\\
		0 & \quad t \geq t_r
	\end{matrix}
	,
\ee
where $t_r > t_0$ is a reference scale (arising as a consequence of dimensional transmutation). Since we are describing a renormalized trajectory, there should be no dependence on the bare scale. 
Moreover, because we want to hit the fixed-point with $\eta_\star = -2$ as $t\rightarrow -\infty$ and the Gaussian fixed-point as $t\rightarrow +\infty$ we conclude that
\be
	w_{t_0}\bigl(p^2 e^{-2(t-t_0)}) = \bigl(a_0 + a_2 p^2 e^{-2t}\bigr) e^{2(t_0-t_r)}
\ee
so that
\be
	h_t(p^2)
	=
	\Biggl\{
	\begin{matrix}
		a_0 p^2 e^{2(t-t_r)} + a_2 e^{-2t_r} p^4 & \quad t<t_r,
	\\
		a_0 p^2 + a_2 p^4 e^{-2t} & \quad t \geq t_r.
	\end{matrix}
\label{eq:RT-jump}
\ee
There are two comments to make.
First, since there is no dependence on the bare scale we can trivially send $t_0 \rightarrow -\infty$ and so now the reference scale simply satisfies $\infty > t_r > -\infty$. Secondly, that a factor of $e^{-t_r}$ remains in the limit $t \rightarrow -\infty$ is inconsequential: it is positive definite and can simply be absorbed into $a_2$.

From this example, it should be apparent that we can in fact choose any $\eta(t)$, so long as it asymptotes smoothly to the following limits:
\be
	\lim_{t \rightarrow -\infty} \eta(t) = -2,
	\qquad
	\lim_{t \rightarrow +\infty} \eta(t) = 0.
\ee
In general, then, the renormalized trajectory between the two fixed-points is described by
\be
	h_t(p^2)
	=
	p^2 \bigl(a_0 + a_2 p^2 e^{-2t}\bigr) 
	e^{-\int^t_{t_r} dt' \eta(t')}
,
\ee
where, as before, $\infty > t_r > -\infty$ is some arbitrary reference scale.

From this analysis, it is clear that we can construct a trajectory between any pair
of fixed-points described above, so long as $\eta_\star^{\mathrm{UV}} < \eta_\star^{\mathrm{IR}}$.
Taking
\be
	\lim_{t\rightarrow -\infty} \eta(t) = -2n
	\qquad
	\lim_{t\rightarrow +\infty} \eta(t) = -2m,
\ee
the solution is given by
\be
	h_t(p^2) = p^2
	\bigl(
		a_{2n} p^{2n} e^{-2nt} + a_{2m} p^{2m} e^{-2mt}
	\bigr)
	e^{-\int^t_{t_r} dt' \eta(t')}
.
\label{eq:RT-General}
\ee
So long as $n>m$ we see that the limits $t \rightarrow \pm \infty$ both exist and both correspond to the anticipated fixed-points.
That all of these flows exhibit $\eta_\star^{\mathrm{UV}} < \eta_\star^{\mathrm{IR}}$ is consistent with the conjecture of  Vicari and Zinn-Justin that the most stable fixed-point is the one with the largest value of $\eta_\star$~\cite{Vicari+Zinn}.

Let us note that the network of flows given by~\eq{RT-General} serves
 as a strong warning for attempting to invoke asymptotic safety scenarios. 
For even restricting ourselves to purely two-point theories we have succeeded in uncovering an infinite number of asymptotically safe 
trajectories (an infinite number of which flow to the Gaussian fixed-point in the IR), all of which are
unphysical.

\section{Asymptotic Safety in Quantum Gravity}

There is a neat argument due to Percacci and Perini~\cite{Percacci-NewtonConstant} that Newton's constant, rendered dimensionless by the effective scale, must stop running at a fixed-point despite playing a seemingly analogous role to $Z$ above. The logic is as follows. Supposing (for argument's sake) that the effective average action possesses the usual Einstein-Hilbert term, plus a cosmological constant (not to be confused with the effective scale, previously denoted by $\Lambda$ but now, following tradition, denote by $k$) we can write, in dimensionful variables,
\be
	\Gamma_k [g_{\mu\nu}]
	=
	\frac{1}{16\pi G(k)}
	\Int{x} \sqrt{g}
	\bigl[
		2 \Lambda(k) - R + \cdots
	\bigr],
\ee
where the ellipsis stands for whatever other interactions there might be. In the linearized theory, $1/G$ multiplies the graviton kinetic term and so plays a similar role to the field strength renormalization in the scalar field theory discussed above. 

Now, it would appear that $G$ is an inessential coupling because it can be removed from the action by a field redefinition \ie\ by a rescaling of the metric. Doing this would produce an $\eta$ term on the \lhs\ of the flow equation, as in~\eq{Gamma-Flow}. But there is a subtlety. Recall that to uncover fixed-points we should transfer to dimensionless variables, as in~\eqs{Rescale-Mom}{Rescale-Fields}. In a gravitational theory, however, this is achieved by redefining the metric. As recognized in~\cite{Percacci-NewtonConstant} we cannot, therefore, both transfer to dimensionless variables \emph{and} remove $G$ from the action. Consequently, $G$ is effectively an essential coupling. Transferring to dimensionless variables causes $G$ to be replaced by
\be
	\tilde{G}(k) = G(k) k^{\D-2}.
\ee 
Since, as just discussed, $\tilde{G}$ must appear in the action it must stop flowing at a fixed-point. Consequently,
\be
	k \der{\tilde{G}_\star}{k} = 0,
	\qquad
	G(k) \sim k^{2-\D}
.
\ee
This result is often rephrased in terms of a so-called anomalous dimension:
\be
	\eta \equiv k \der{\ln G(k)}{k},
\ee
and so we recover the result of Lauscher and Reuter~\cite{Reuter+Lauscher-HD} that, for any fixed-point besides the Gaussian one, $\eta_\star = 2-\D$.

Before moving on let us note that, in this context, the term `anomalous dimension' is really a misnomer, since there is nothing anomalous about it! We have determined $\eta_\star$ purely by scaling arguments. Contrast this to the scalar field theory discussed above where, because $Z$ can be removed from the action even when working in dimensionless variables, there is no need for $Z$ to stop running at what is, for the other couplings, a fixed-point. In this case, $\eta$ is not determined by scaling arguments, as discussed above. 

However, this is not the end of story. To actually construct a flow equation for quantum gravity requires that we introduce a coarse-graining procedure. However, this is conceptually non-trivial since we do not know the metric ahead of time and we need a metric to define distances. The solution advocated by Reuter~\cite{Reuter-Genesis,Reuter-BiMetricQEG} is to introduce an \emph{unspecified} background metric, $\bar{g}_{\mu\nu}$. In the context of the effective average action approach we thus have that
\be
	g_{\mu\nu} = \bar{g}_{\mu\nu}  + h_{\mu\nu},
\ee
where 
$h_{\mu\nu}$ is a fluctuation which need not be small. Whilst background invariance is sacrificed in this approach, background covariance is preserved by leaving $\bar{g}_{\mu\nu}$ undetermined. A cutoff can now be introduced by using the background metric. In practice, this amounts to modifying the two-point term for $h_{\mu\nu}$ (and the ghosts) with a kernel that depends on $\bar{g}_{\mu\nu}$. The upshot of this is that the effective average action depends \emph{separately} on $\bar{g}_{\mu\nu}$ and $h_{\mu\nu}$. 

With this in mind, let us return to the two issues of transferring to dimensionless variables and performing field redefinitions. Because $\bar{g}_{\mu\nu}$ is used to define distances, it follows that the former step can be achieved by redefining this field. But since the action depends separately on $h_{\mu\nu}$, there is nothing to stop us from independently redefining this field. Consequently, the coupling in front of the two-point kinetic term for $h_{\mu\nu}$ is, just as in our scalar example, inessential. Therefore, the anomalous dimension (in the true sense of the word) of the fluctuation field does not follow from scaling arguments alone and so must be computed using the flow equation. (Note that this has already been recognized to be the case for the gauge-fixing ghosts, whose anomalous dimension has, within a certain truncation scheme, already been computed~\cite{Eichhorn-Ghost}).

What does this imply in the context of calculations that have been done to date? Until recently, all computations have been performed within truncation schemes whereby the action depends only on a single metric. Such truncations are insensitive to the above considerations and so any fixed-point will necessarily have an `anomalous dimension' of $2-\D$. In~\cite{Reuter-BiMetricQEG}, however, a so-called `bi-metric' truncation was used, in which the action was taken to be a functional of both $g_{\mu\nu}$ and $\bar{g}_{\mu\nu}$. Nevertheless, the truncation was of a form such that the $h_{\mu\nu}$ two-point piece renormalizes in the same way as the $\sqrt{\bar{g}} R(\bar{g})$ term. Thus, although two versions of Newton's constant were defined---one appropriate to $\sqrt{\bar{g}} R(\bar{g})$ and one appropriate to $\sqrt{{g}} R({g})$---both are compelled to come with an anomalous dimension of $2-\D$. Consequently, the possibility that $h_{\mu\nu}$ is associated with a genuinely anomalous dimension, say $\eta_\star^h$, has so far not been investigated.

Should such investigations be pursued, there is the possibility of finding a richer spectrum of fixed-points than have been found to date. On the one hand, there is nothing, \emph{a priori}, which prohibits `non-anomalous'  fixed-points with $\eta_\star^h = 2-\D$ (just as there is nothing which prohibits $\eta_\star=0$ in the scalar case). Indeed, since two such fixed-points are found in the `bi-metric' truncation of~\cite{Reuter-BiMetricQEG} it is reasonable to suppose that the existence of such fixed-points is not simply an artefact of single-metric truncations. On the other hand, it might be the case that fixed-points exist with $\eta_\star^h \neq 2-\D$. Either way, particularly given the lessons learned above from scalar field theory, it is crucial to determine whether or not any given non-trivial fixed-point theory is unitary. Quite apart from truncation issues, this is presumably a difficult question to answer not least because even if the exact fixed-point solution were known, the metric would still have to be determined from the effective Einstein equations (and there is no guarantee, \emph{a priori}, that flat space would be a solution).

\section{Conclusion}

It has been emphasised that the flow equations for the effective average action---both in the case where the redundant field strength renormalization is left in the action and the case where it is scaled out---are related to Wilsonian flows via Legendre transforms. The effective average action formalism makes it particularly easy to exactly compute an infinite set of asymptotically safe trajectories which, though in a sense trivial, serve to illustrate the pitfall of conflating unitarity with renormalizability (even when the latter holds nonperturbatively). With this in mind, it is argued that more complex truncations than have been attempted to date allow for the possibility of a new class of fixed-points in asymptotically safe quantum gravity. This could potentially be relevant to the issue of unitarity. It might also allow for a resolution of the disagreement between the spectral dimension computed using the Exact RG~\cite{Reuter-Fractal} and the recent lattice result~\cite{Lattice-AS-Evidence}. 

\begin{acknowledgments}
	I would like to thank Alessandro Codello, Roberto Percacci and Martin Reuter for
	correspondence. Particular thanks to Hugh Osborn for providing me with a draft
	of~\cite{HO-Remarks}, for bringing~\cite{Vicari+Zinn} to my attention and for useful
	comments on the manuscript.
	This work was supported by the Science and Technology Facilities Council 
	[grant number ST/F008848/1].
\end{acknowledgments}

\appendix

\section{Details}
\label{app:Details}

To show that the flow equation for $\Gamma$, \eq{Gamma-Flow}, does indeed follow from
the modified Polchinski equation, \eq{Ball},
given~\eq{P}, \eqs{Q}{R}, let us start by noting that the analogue of~\eq{flows} reads:
\begin{multline}
	\flow \bigr\vert_\field \Sint_\Lambda[\field] =
	\flow \bigr\vert_\Phi \Gamma_\Lambda[\Phi]
\\ 
	+ \field \cdot \dot{\mathcal{P}}^{\Lambda_0}_\Lambda \cdot \Phi
	- \hf \field \cdot \dot{\mathcal{Q}}^{\Lambda_0}_\Lambda \cdot \field 
	- \hf \Phi \cdot \dot{\mathcal{R}}^{\Lambda_0}_\Lambda \cdot \Phi.
\end{multline}
Using~\eq{Phi-def} together with
\be
	\fder{\Gamma_\Lambda[\Phi]}{\Phi(p)} = \mathcal{P}_\Lambda^{\Lambda_0}(p^2) \field(-p) 
	- \mathcal{R}_\Lambda^{\Lambda_0}(p^2) \Phi(-p)
\ee
it is straightforward to show that~\eq{Gamma-Flow}---with $f$ and $F$ given by~\eq{f+F}---is implied by~\eq{Ball} (up to a discarded vacuum energy term), so long as we take $\mathcal{P}$, $\mathcal{Q}$ and $\mathcal{R}$ to satisfy:
\begin{subequations}
\begin{align}
	\Bigl(\totalflow{} - \eta\Bigr) \mathcal{P} & = -\mathcal{P} \mathcal{Q} \dd,
\label{eq:P-flow}
\\
	\Bigl(\totalflow{} - \eta\Bigr) \mathcal{Q} & = \eta \ep^{-1} - \mathcal{Q}^2 \dd,
\label{eq:Q-flow}
\\
	\Bigl(\totalflow{} - \eta\Bigr) \mathcal{R} & = - \mathcal{P}^2 \dd.
\label{eq:R-flow}
\end{align}
\end{subequations}

Focusing first on~\eq{Q-flow}, we follow~\cite{HO-Remarks} and define
\be
	\tilde{\mathcal{Q}} \equiv e^{\zeta} \bigl(\mathcal{Q} + \ep^{-1} \bigr),
\ee
which leads to
\be
	\totalflow{\tilde{\mathcal{Q}}} = -e^{-\zeta} \dd \tilde{\mathcal{Q}}
	\bigl(
		\tilde{\mathcal{Q}} - 2 \ep^{-1} e^{\zeta}
	\bigr).
\ee
It is straightforward to check that this implies
\be
	\totalflow{} \frac{1}{\tilde{\mathcal{Q}} \ep^2} = e^{-\zeta} \totalflow{} \frac{1}{\ep}
\ee
and, integrating up, we find~\eq{Q}.

Next, let us observe from~\eq{sigma} that
\be
	\totalflow{\sigma} = \frac{\dot{\cutoff}}{\cutoff}\bigl(1-\sigma \bigr) - \eta \sigma.
\label{eq:sigma-flow}
\ee
Utilizing this result, we can see from~\eq{Q} that
\be
	\mathcal{Q} \dd = \frac{\dot{\cutoff}}{\cutoff} \Bigl(\frac{1}{\sigma} -1  \Bigr)
	= 
	\frac{1}{\sigma}\totalflow{\sigma} + \eta
	;
\ee
substituting this into~\eq{P-flow}, \eq{P} immediately follows.
Again utilizing~\eq{sigma-flow}, it is apparent that
\be
	-\mathcal{P}^2 \dd =  b^2 p^2 e^{-\zeta} \totalflow{} \frac{\cutoff e^\zeta}{\sigma}.
\ee
Substituting this into~\eq{R-flow}, it is simple to show~\eq{R}.

\bibliography{../../Biblios/ERG.bib,../../Biblios/Books.bib,../../Biblios/Foundations,../../Biblios/Misc}

\end{document}